\documentclass[a4paper,12pt]{article}
\pdfoutput=1
\usepackage{graphicx, rotating}
\usepackage{hyperref}
\usepackage{slashed}
\usepackage{ifpdf}

\ifx\pdfoutput\undefined
\usepackage[dvips,bookmarks=false]{hyperref}	
\else
\usepackage{hyperref}	
\fi
\hypersetup{colorlinks,bookmarksopen,bookmarksnumbered,citecolor=verdes,
linkcolor=blus,pdfstartview=FitH,urlcolor=rossos}
\def\hhref#1{\href{http://arxiv.org/abs/#1}{#1}} 

\usepackage{amsfonts}
\usepackage{amsmath}
\usepackage{slashed}

\newcommand{\beq}{\begin{equation}}
\newcommand{\eeq}{\end{equation}}
\newcommand{\fig}[1]{~\ref{fig:#1}}

\newcount\Mac  \Mac=1  
\newcommand{\ifMac}[2]{\ifnum\Mac=1 #1 \else #2 \fi}
\def\putps(#1,#2)(#3,#4)#5#6{\ifnum\Mac=1 \put(#1,#2){\special{picture #5}}
\else  \put(#3,#4){\includegraphics{#6}} \fi}

\newcommand{\One}{\hbox{1\kern-.24em I}}

\newcommand{\GeV}{\,{\rm GeV}}

\newcommand{\eq}[1]{~{\rm (\ref{eq:#1})}}

\newcommand{\lascia}[1]{}
\makeatletter
\def\art{\@ifnextchar[{\eart}{\oart}}
\def\eart[#1]#2#3#4#5#6{{\rm #2}, {#3 #4} {\rm (#6) #5} [arXiv:{\hhref{#1}}]}
\def\hepart[#1]#2{{\rm #2, arXiv:\hhref{#1}}}
\newcommand{\oart}[5]{{\rm #1}, {#2 #3} {\rm (#5) #4}}

\newcounter{alphaequation}[equation]
\def\thealphaequation{\theequation\hbox to
0.6em{\hfil\alph{alphaequation}\hfil}}
\def\eqnsystem#1{
\def\@eqnnum{{\rm (\thealphaequation)}}
\def\@@eqncr{\let\@tempa\relax \ifcase\@eqcnt \def\@tempa{& & &} \or
  \def\@tempa{& &}\or \def\@tempa{&}\fi\@tempa
  \if@eqnsw\@eqnnum\refstepcounter{alphaequation}\fi
\global\@eqnswtrue\global\@eqcnt=0\cr}
\refstepcounter{equation} \let\@currentlabel\theequation \def\@tempb{#1}
\ifx\@tempb\empty\else\label{#1}\fi
\refstepcounter{alphaequation}
\let\@currentlabel\thealphaequation
\global\@eqnswtrue\global\@eqcnt=0 \tabskip\@centering\let\\=\@eqncr
$$\halign to \displaywidth\bgroup \@eqnsel\hskip\@centering
$\displaystyle\tabskip\z@{##}$&\global\@eqcnt\@ne
\hskip2\arraycolsep\hfil${##}$\hfil& \global\@eqcnt\tw@\hskip2\arraycolsep
$\displaystyle\tabskip\z@{##}$\hfil
\tabskip\@centering&\llap{##}\tabskip\z@\cr}

\def\endeqnsystem{\@@eqncr\egroup$$\global\@ignoretrue} \makeatother

\def\Lag{{\cal L}}

\def\circa#1{\,\raise.3ex\hbox{$#1$\kern-.75em\lower1ex\hbox{$\sim$}}\,}

\usepackage{multicol}
\usepackage{color}
\definecolor{rosso}{cmyk}{0,1,1,0.4}
\definecolor{rossos}{cmyk}{0,1,1,0.55}
\definecolor{rossoc}{cmyk}{0,1,1,0.2}
\definecolor{blu}{cmyk}{1,1,0,0.3}
\definecolor{blus}{cmyk}{1,1,0,0.6}
\definecolor{bluc}{cmyk}{1,1,0,0.1}
\definecolor{verde}{cmyk}{0.92,0,0.59,0.25}
\definecolor{verdec}{cmyk}{0.92,0,0.59,0.15}
\definecolor{verdes}{cmyk}{0.92,0,0.59,0.4}
\definecolor{grigio}{cmyk}{0,0,0,0.07}
\definecolor{rosa}{cmyk}{0,0.1,0.1,0.02}
\definecolor{rosino}{cmyk}{0,0.05,0.05,0.02}
\definecolor{rosas}{cmyk}{0,0.3,0.25,0.05}
\definecolor{celeste}{cmyk}{0.1,0,0,0.02}
\definecolor{giallino}{cmyk}{0,0,0.4,0.02}
\definecolor{rosso}{cmyk}{0,1,1,0.4}
\definecolor{rossos}{cmyk}{0,1,1,0.55}
\definecolor{rossoc}{cmyk}{0,1,1,0.2}
\definecolor{blu}{cmyk}{1,1,0,0.3}
\definecolor{bluc}{cmyk}{1,1,0,0.1}
\definecolor{blucc}{cmyk}{0.7,0.5,0,0}
\definecolor{viola}{cmyk}{0,1,0,0.6}
\definecolor{viola2}{cmyk}{0,1,0.2,0.6}
\definecolor{verde}{cmyk}{0.92,0,0.59,0.25}
\definecolor{verdec}{cmyk}{0.92,0,0.59,0.15}
\definecolor{verdes}{cmyk}{0.92,0,0.59,0.4}
\definecolor{verdino}{cmyk}{0.12,0,0.09,0.05}
\definecolor{giallo}{cmyk}{0,0,1,0}
\definecolor{gialloverde}{cmyk}{0.44,0,0.74,0}

\font\tenrsfs=rsfs10 at 12pt

\font\sevenrsfs=rsfs7
\font\fiversfs=rsfs5
\newfam\rsfsfam
\textfont\rsfsfam=\tenrsfs
\scriptfont\rsfsfam=\sevenrsfs
\scriptscriptfont\rsfsfam=\fiversfs
\def\mathscr#1{{\fam\rsfsfam\relax#1}}
\def\Lag{\mathscr{L}}

\begin{document}
\color{black}
\vspace{0.5cm}
\begin{center}
{\Huge\bf\color{black} Hints for a non-standard \\[5mm]   Higgs boson from the LHC}
\bigskip\color{black}\vspace{0.6cm} \\[3mm]
{{\large\bf  Martti Raidal$^{a}$ and  Alessandro Strumia$^{a,b}$}
} \\[7mm]
{\it  (a) National Institute of Chemical Physics and Biophysics, Ravala 10, Tallinn, Estonia}\\[3mm]
{\it  (b) Dipartimento di Fisica dell'Universit{\`a} di Pisa and INFN, Italia}\\[3mm]
\end{center}
\bigskip

\centerline{\large\bf\color{blus} Abstract}
\begin{quote} 
\color{black} 
We  reconsider Higgs boson invisible decays into Dark Matter in the light of  recent Higgs searches at the LHC.
Present hints in the CMS and ATLAS data favor a non-standard Higgs boson with approximately $50\%$ invisible branching ratio, 
and mass around 143 GeV.  This situation can be realized within the simplest thermal scalar singlet Dark Matter model, predicting a
Dark Matter mass  around 50 GeV and direct detection cross section just below present bound.
The present runs of the {\sc Xenon}100 and LHC experiments can test this possibility.
\end{quote}

\subsection*{Introduction}
Both  the CMS~\cite{CMSLP} and ATLAS~\cite{ATLASLP} experiments have presented their combined searches for the standard model (SM)
Higgs boson, based on luminosities  between 1 fb$^{-1}$ and 2.3 fb$^{-1}$ depending on the search channel. In the light Higgs mass region
preferred by precision data, both experiments exclude the SM Higgs boson with mass greater than 146~GeV. 

\medskip

At the same time, both experiments find the following  hints in their data:
\begin{itemize}
\item In the most sensitive channel $h \to WW^*\to 2\ell \, 2\nu$ they find a broad excess of events over the computed background.  In view of the undetected neutrinos this channel poorly allows to reconstruct the
Higgs boson mass and any $m_h$ between 100 and $2M_W$ can reproduce the apparent excess roughly equally well.
\item Furthermore, the ``golden" (but less sensitive) channel $h\to ZZ^*$ shows an excess of signal events for
the Higgs boson mass around 143~GeV. These features appear consistently in both experiments. 
(Other higgs masses are favored by the single experiments; but only this value is favored by both).

\end{itemize}

\begin{figure}[t]
$$\includegraphics[width=0.5\textwidth]{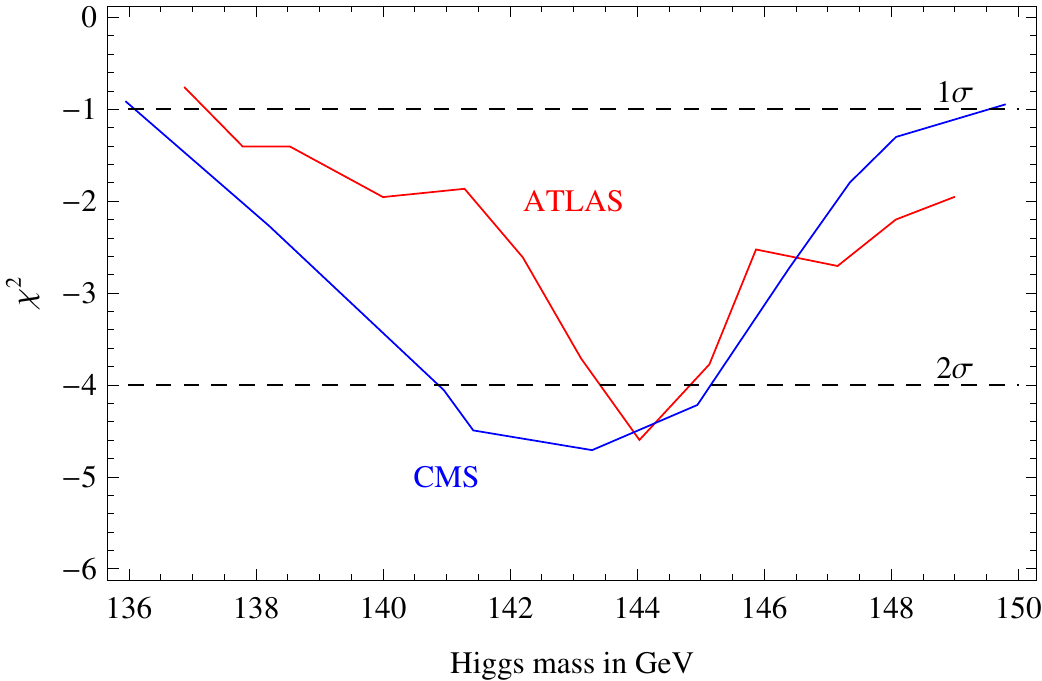}$$
\caption{\em Hints in CMS and ATLAS data for a Higgs boson with mass around 143 GeV.
The figure is obtained converting into a $\chi^2$ the experimental fits reported in terms of probability values in Refs.~\cite{CMSLP,ATLASLP}.
\label{fig:hints}}
\end{figure}

Needless to say, the significance of such features is statistically inconclusive: slightly more than $2\sigma$ in each experiment,
as shown in fig.\fig{hints}.
The significance will get  reduced after taking into account ``look elsewhere effects", i.e.\ the fact that 
because of statistical fluctuations a fake peak can appear at many different $m_h$. 

The best fit value of the SM Higgs boson mass is around
120~GeV~\cite{CMSLP}, in the region where the experiments do not yet have sensitivity to discover it.
The SM Higgs boson with mass $m_h$ around 143~GeV is not favored by the data, because it
would have given an excess in $h\to WW^*$ about twice larger than the observed hint.

While these hints could well be statistical fluctuations or systematical artifacts related to the experiments, 
the purpose of this note is to point out a more exciting possibility: both hints for $h\to WW^*$ and $h\to ZZ^*$ can be jointly
fitted by a non-standard Higgs boson.

Various possibilities exists: new particles at the weak scale can  i)  reduce the gluon fusion Higgs production cross section;
ii)  increase the Higgs BR into $b\bar b$, which is difficult to see experimentally, at the expense of branching ratio into $WW^*$ and $ZZ^*$;
iii) add an invisible  branching ratio, correspondingly reducing all SM branching ratios.
If the $WW^*$ and $ZZ^*$ Higgs rates are reduced by $0.5\pm0.2$, the interpretation of LHC Higgs boson searches changes and  $m_h\approx 143$~GeV
becomes the best fit value of the Higgs boson mass supported by the data. 
Thus the LHC data indicates to a ``physicist dream" scenario in which Higgs boson couples equally strongly
to the SM particles and to the unknown new physics.

Possibility iii) is easily realized compatibly with all present negative searches for new physics:
invisible Higgs boson decays~\cite{Eboli:2000ze}  can occur in many theories beyond the SM including 
supersymmetry~\cite{Griest:1987qv}, extra dimensions~\cite{Datta:2004jg}, 
Majoron models~\cite{Romao:1992zx}, 
fourth generation~\cite{Khlopov},
models with non-trivial hidden sectors~\cite{Schabinger:2005ei}, 
``hidden valley" models~\cite{Strassler:2006im},  models with scalar Dark Matter~\cite{McDonald,Barbieri:2006dq,Kadastik:2009dj}  etc.
Model independently,  LHC needs about 30 fb$^{-1}$ data  to probe at 95\% C.L. a 
$\approx 50\%$ Higgs branching ratio into invisible channels for $m_h\approx 140$~GeV~\cite{prospects}. Thus the interpretation proposed in this work
can be tested by the LHC alone. 

Among the large number of  new physics scenarios the most motivated are the ones
predicting Dark Matter (DM). Indeed, the existence of DM~\cite{WMAP7}  is presently the only firm evidence of new physics.
If invisible Higgs decays are observed at the LHC, it is most natural to assume that the Higgs boson decays into DM.
In this case the LHC probes both the mass and the couplings of the DM particles and predicts the DM-nucleon scattering cross section
probed by direct detection experiments. 
The latter, such as {\sc Xenon}100, can cross check the LHC results.  

In the light of tight constraints
on supersymmetric DM coming from null results from the LHC~\cite{CMS, ATLAS} and {\sc Xenon}100~\cite{Xenon100},
the global fit~\cite{Farina:2011bh} now excludes $M_{\rm DM}<m_h/2$ in the CMSSM.
In order to get an invisible Higgs width we focus on the most economical DM model, extending the SM with a real singlet scalar $S$~\cite{McDonald}.
The capability of {\sc Xenon}100 and the LHC to constrain this model was recently studied  in Refs.~\cite{Farina:2011bh,Mambrini}.
In this letter we show that the scalar singlet DM model~\cite{McDonald} is consistent with the 
LHC and DM direct detection data and predicts DM with mass $M_{\rm DM}\approx 50$~GeV to be discovered by the {\sc Xenon}100 experiment.

\subsection*{The Dark Matter interpretation}
We consider the simplest DM model  obtained adding to the SM a  real singlet scalar field $S$
coupled to the Higgs doublet $H$ as described by the following Lagrangian~\cite{McDonald}:
\beq \Lag =\Lag_{\rm SM } + \frac{(\partial_\mu S)^2}{2} - \frac{m^2}{2}S^2
-\lambda S^2 |H|^2 - \frac{\lambda_S}{4!} S^4\  .
\label{eq:L}\eeq
 Thanks to the discrete symmetry $S\to -S$ the singlet $S$ becomes a good DM candidate.
 For simplicity and minimality we assume $S$ to be real.
Given that $\lambda_S$ is essentially irrelevant for particle physics phenomenology, the model has only 2 new free parameters: the DM mass 
given by $M_{\rm DM}^2 = m^2 +\lambda V^2$ with $V =246\GeV$, and 
the DM/Higgs coupling $\lambda$. 
The latter  can be fixed assuming that the relic DM abundance equals to its cosmologically measured value.

\begin{figure}[t]
$$\includegraphics[width=0.45\textwidth]{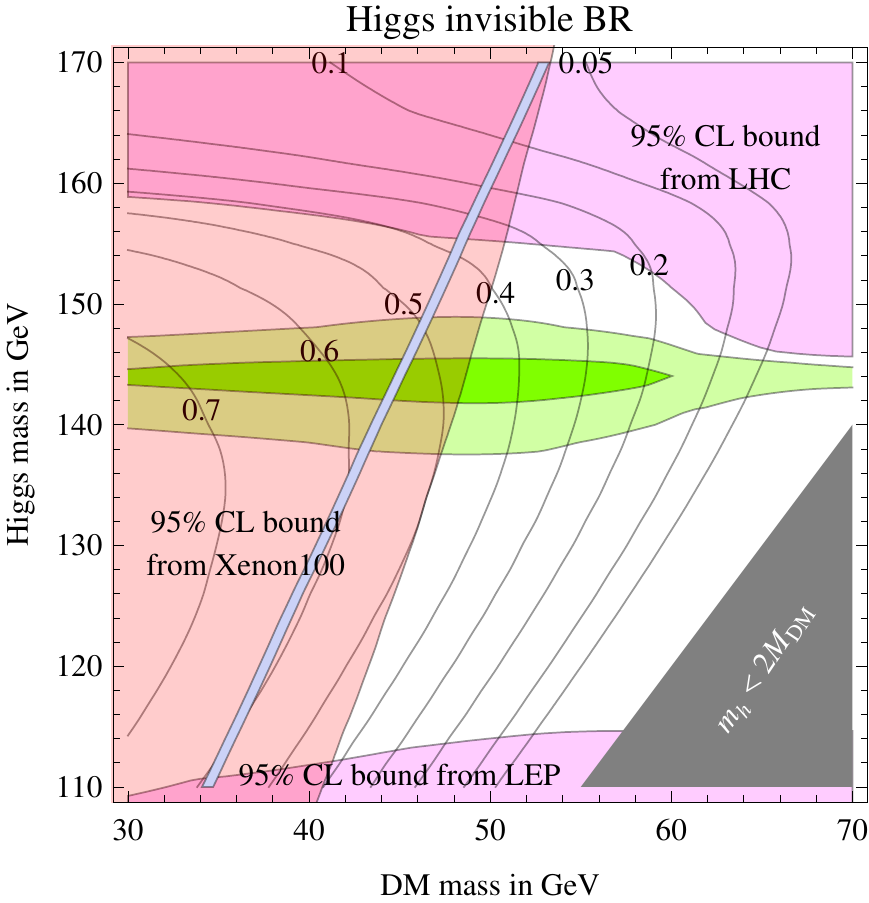}\qquad\includegraphics[width=0.472\textwidth]{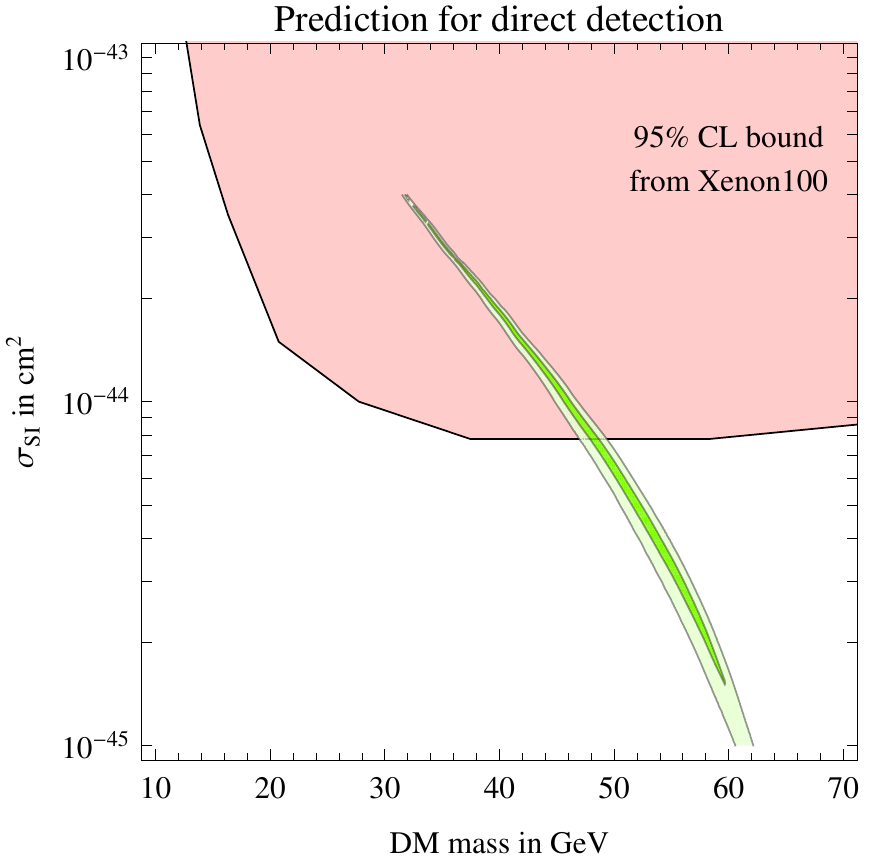}$$
\caption{\em {\bf Left}: Iso-lines of the invisible Higgs boson BR in the scalar DM singlet model, fixing the Higgs/DM coupling from the cosmological
DM abundance.  The green region shows the region favored at $68\%$ and $95\%$ C.L.\ from our estimate of LHC data,
ignoring look-elsewhere-effects.  The red region is excluded by {\sc Xenon}100, and the magenta region is excluded by LHC (upper)
and LEP (lower) higgs searches.
The gray line corresponds to the most minimal model with $m^2=0$.
{\bf Right}: Predicted  DM spin independent direct detection
cross section as a function of the DM mass (green region).
\label{fig:singlet}}
\end{figure}

In this work we compute the DM relic density following~\cite{CDMSth} (that included the 3 body final states whose relevance was emphasized in~\cite{3})
and require it to be equal to the observed value, $\Omega_{\rm DM}=0.112 \pm 0.0056$~\cite{WMAP7}.
As a consequence the model predicts the  Higgs boson invisible decay width 
\beq \Gamma  (h\to SS) = \frac{\lambda^2V^2}{8\pi m_h}\sqrt{1-4\frac{M_{\rm DM}^2}{m_h^2}}\label{eq:GammaSS}\ .\eeq
The corresponding invisible branching ratio is shown by the isocurves in the left panel of fig.\fig{singlet},
as functions of the DM and Higgs boson masses. The  region shaded in red is excluded by {\sc Xenon}100, and the region
shaded in gray has $m_h < 2 M_{\rm DM}$ and consequently no invisible Higgs width.

Performing a na\"{\i}ve combination of the ATLAS~\cite{ATLASLP} and CMS~\cite{CMSLP} hints (by just adding their $\chi^2$ plotted in fig.\fig{hints}),
the region favored at 68\% and 95\% C.L.\ (2 dof) is shown in green.
In this case the Higgs boson mass is predicted to be in a narrow region $m_h=(138-148)$~GeV at 95\% CL. 
The Higgs boson invisible decay branching ratio is preferably in between 0.2 and 0.5.

Fig.\fig{singlet}a extends to vastly different Higgs boson masses, thereby covering
the possibility that the excess in the most sensitive $h\to WW^*$ channel is real, 
while the Higgs boson mass favored by $h\to ZZ^*$ are statistical fluctuations.
In this case the partially invisible Higgs boson can even have a mass below the LEP bound $m_h>115\GeV$,
and its discovery at the LHC requires more luminosity (and work) than expected in case of the SM Higgs boson.

The diagonal gray line in fig.\fig{singlet}a corresponds to the most minimal model $m=0$~\cite{Kadastik:2009dj,CDMSth}, where DM obtains its mass entirely from the electroweak symmetry breaking.
This possibility has now been excluded or disfavored by {\sc Xenon}100, as discussed in~\cite{Farina:2011bh}.
The allowed part of the parameter space corresponds to $m^2>0$.


\medskip

In the right panel of  fig.\fig{singlet} we plot our prediction for the DM spin independent cross section as a function of the DM mass.
As in the left panel, the shaded red region in the plot is excluded at more than $95\%$ C.L.\ (1 dof) by the {\sc Xenon}100 search for direct DM detection.
The Higgs/DM coupling $\lambda$ determines both the Higgs invisible width and
the spin-independent DM direct detection cross section,
\beq\label{eq:f} \sigma_{\rm SI}
= \frac{\lambda^2 m_N^4 f^2}{\pi M_{\rm DM}^2 m_h^4},\ \eeq
where $f$ parameterizes the nucleon matrix element,
\beq \langle{N} | m_q \bar{q}q | N\rangle\equiv  f_q m_N [\bar N N] ,\qquad
f = \sum_{q=\{u,d,s,c,b,t\}}\!\!\!\!\! f_q = \frac{2}{9}+\frac{5}{9}\sum_{q=\{u,d,s\}}\!\! f_q.  \label{eq:f} \eeq
We here assumed $f= 0.30$, based on the lattice result
$f= 0.30\pm0.015$~\cite{lattice}. 
The main uncertainty on $f$
comes from the strange quark contribution $f_s$ in\eq{f}, and other computations give higher more uncertain values, such as 
$f= 0.56\pm 0.11$~\cite{Ellis}. The LHC~\cite{CMSLP,ATLASLP} and LEP~\cite{Barate:2003sz} excluded region are also shown in magneta.

The {\sc Xenon}100 exclusion bound have been plotted assuming  the local DM density $\rho_\odot = 0.3\GeV/{\rm cm}^3$.
This is the canonical value routinely adopted in the literature, with a typical associated error bar of $\pm 0.1$ GeV/cm$^{3}$. 
Recent estimations  found a higher central value closer to $0.4\GeV/{\rm cm}^3$~\cite{Nesti}
that would imply stronger bounds on the cross section $\sigma_{\rm SI}$. 
The DAMA~\cite{DMsig}, {\sc CoGeNT}~\cite{cogent} and CRESST-II~\cite{cresst} direct DM searches claim positive hints, that are difficult to
reconcile with bounds from other experiments such as {\sc Xenon} even assuming highly non-standard DM models~\cite{DMfits} --- an avenue that we do not explore here.

\medskip

We notice that expressions similar to eq.\eq{GammaSS} apply in all models where Higgs decays to DM, such that a similar value of $\lambda$
is obtained: consequently a similar detectable DM direct detection is predicted in more generic models.

\bigskip

\subsection*{Conclusions}
Motivated by the existence of DM, we have studied the implications of Higgs boson invisible decays into DM for  the recent CMS and ATLAS 
Higgs  searches. We find that the best fit of the present data is provided in terms of a non-standard Higgs boson, 
with a $\approx 50\%$ invisible branching ratio and possibly a mass around 143 GeV. 
We demonstrated that the simplest scalar singlet model for DM can provide such an invisible width. 
The needed Higgs boson/DM coupling, $\lambda\approx 0.03,$ is consistent with DM as a thermal relic 
and predicts the direct detection cross section just below present bounds. 
The present runs of the {\sc Xenon}100 and LHC experiments can test this possibility.

\paragraph{Acknowledgements}
We thank G. Giudice, A. Korytov and M. Papucci for discussions. This work was performed in the CERN TH-LPCC summer institute on LHC physics. 
This work was supported by the ESF grants  8090, 8499, MTT8 and by SF0690030s09 project.


\end{document}